\documentclass[final,3p,twocolumn]{elsarticle}

\makeatletter
\def\ps@pprintTitle{%
  \let\@oddhead\@empty
  \let\@evenhead\@empty
  \let\@oddfoot\@empty
  \let\@evenfoot\@oddfoot
}
\makeatother

\usepackage[pagewise]{lineno}
\modulolinenumbers[1]
\usepackage{hyperref}
\usepackage{float}
\usepackage{svg}
\hypersetup{colorlinks=true,pdfborder=0 0 0}
\usepackage{subcaption}
\usepackage{amsmath,bm,siunitx,amssymb,caption,booktabs}
\usepackage{ulem}

\usepackage{xspace}

\usepackage{textcomp,gensymb}
\usepackage{cleveref}
\usepackage{ulem}
\usepackage{xcolor}

\journal{arXiv}

\begin{document}

\begin{frontmatter}

\title{Lattice genome: representation and analysis of heterogeneous crystalline microstructures}

\address[az-mse]{Department of Materials Science and Engineering, University of Arizona, Tucson, AZ 85721, USA}
\address[uiuc]{Materials Science and Engineering Department, University of Illinois Urbana-Champaign, Urbana, IL, USA}
\address[az-am]{Graduate Interdisciplinary Program in Applied Mathematics, University of Arizona, Tucson, AZ 85721, USA}

\author[az-mse]{Jiayang Wang}
\author[uiuc]{Mathieu Calvat}
\author[uiuc]{J.C. Stinville}
\author[az-mse,az-am]{Marat I. Latypov\corref{cor1}}
\cortext[cor1]{corresponding author}
\ead{latmarat@arizona.edu}

\begin{abstract}

Inspired by the concept of a generalized materials genome, we introduce the notions of \textit{lattice gene} and \textit{lattice genome} for crystalline materials. A lattice gene is a compact representation of the local crystalline structure obtained by encoding the Kikuchi diffraction patterns with a variational autoencoder.  We show that this representation satisfies key criteria for a materials gene: compactness, experimental accessibility, existence of a distance metric reflecting structural similarity, and sufficient information content for reconstructing the original diffraction patterns. The lattice genome is the spatially resolved collection of lattice genes across a representative area mapped by electron backscatter diffraction (EBSD), which captures mesoscale heterogeneity that ultimately controls properties. We demonstrate three applications of the lattice genome: (i) latent component maps that visualize grain-scale and intra-grain heterogeneities, (ii) domain segmentation based on distance and angle metrics in the latent space, and (iii) kernel and domain latent vector spreads that quantify intragranular heterogeneity as high-dimensional analogs of kernel average misorientation and grain orientation spread. All three tools are validated on microstructures of additively manufactured and wrought Ni-base superalloys in as-built and recrystallized conditions.

\end{abstract}


\end{frontmatter}

\section{Introduction}

The Materials Genome Initiative (MGI) \cite{MGI} was founded on a biological analogy that materials, like organisms, possess an underlying code that dictates their behavior. Yet, fifteen years later \cite{depablo2019new}, there is surprisingly little clarity on what "materials genes" could actually look like in practice. Billinge recently pondered whether materials have a genome at all \cite{billinge2024materials}. While the answer is negative in the strictly biological sense, the author puts forward a notion of generalized genome that \textit{can} apply to materials. Based on parallels with DNA/RNA, a generalized gene is a vector that encodes the 3D arrangement of atoms. With this notion, desirable characteristics for a genetic code of materials include compactness, uniqueness, accessibility from experiments, and the existence of a distance metric between two codes that represents the similarity between the underlying materials structures. Billinge proposes atomic pair distribution functions (PDF) and graphs as promising descriptions satisfying these criteria \cite{billinge2024materials}. PDF provides a powerful atomic-scale description of molecules, defect-free single crystals, and other macroscopically uniform materials, but lacks the spatial awareness needed to describe heterogeneous materials at the mesoscale. 

Spatial heterogeneity is a defining characteristic of crystalline materials, such as structural alloys, where non-uniform distributions of the lattice (phases), lattice orientation (grains), and lattice defects govern macroscale properties and performance \cite{adams2012microstructure, kalidindi2015hierarchical}. The notion of materials genes for spatially non-uniform materials at the mesoscale has not been explored. Addressing this gap requires a combination of a characterization approach that can spatially resolve microstructural heterogeneity and encoding of the data to efficiently carry the "genetic" information of the lattice.

Electron backscatter diffraction (EBSD) is a powerful technique that addresses the spatial characterization need \cite{schwartz2009electron}. EBSD maps Kikuchi diffraction patterns across large areas at the mesoscopic length scale with significant automation and speed \cite{adams1993orientation}. In conventional EBSD workflows, these Kikuchi patterns are analyzed to extract specific physical quantities such as crystal structure \cite{michael1992crystallographic, nowell2004phase}, crystallographic orientation \cite{wright2015introduction,chen2015dictionary}, lattice rotation \cite{wilkinson2006high}, lattice expansion \cite{wilkinson2006high}, geometrically necessary dislocation densities \cite{pantleon2008resolving}, and total dislocation densities \cite{wang2023dislocation}. While physically intuitive, these descriptors each capture only a single, isolated aspect of the lattice. The rich information content of a Kikuchi pattern is thus reduced to individual and inherently lossy descriptors. Even combining multiple descriptors does not recover all the information embedded in the diffraction data \cite{calvat2025learning}.

This information loss motivated new data-driven representations that can capture more or even all of the information in Kikuchi patterns in a compact form. Early efforts used multivariate statistical methods such as principal component analysis (PCA) \cite{wilkinson2019applications, bonnet1998multivariate, brewer2008multivariate, wright2015electron} or matrix factorization \cite{mcauliffe2020spherical,chauniyal2024employing}, which can produce low-dimensional representations enabling microstructure segmentation \cite{wilkinson2019applications,chauniyal2024employing} or accelerated indexing \cite{varley2026accelerating}. More recently, deep neural networks have been applied to Kikuchi patterns for tasks including crystal symmetry determination \cite{kaufmann2020crystal}, indexing \cite{ding2020indexing}, segmentation \cite{mcauliffe2020spherical}, and deformation state analysis \cite{lu2023crystal}. However, these deep learning efforts have mostly focused on aiding the extraction of \textit{known} physical features from diffraction data, rather than producing a holistic and compact encoding of all the information from a Kikuchi pattern.

Variational autoencoders (VAEs) \cite{kingma2013auto} offer a path toward such holistic encoding \cite{vizoso2025decoding}. A VAE trained on Kikuchi patterns learns to compress each pattern into a low-dimensional latent vector from which the original pattern can be reconstructed to a pre-defined accuracy \cite{liu2025learning,calvat2025learning}. Since a decoder is trained in tandem with the encoder, the latent vector retains sufficient information to reconstruct the full diffraction pattern containing all of the lattice information encoded in the pattern, rather than individual descriptors (\textit{e.g.}, orientation or dislocation content). Calvat \textit{et al.}\ \cite{calvat2025learning} recently demonstrated this approach, showing that spatial mapping of latent features from a VAE encoder reveals microstructural heterogeneities (including dislocation cell structures and lattice rotation domains) with sensitivity exceeding conventional EBSD analysis methods.

In this paper, we argue that such latent representations of Kikuchi patterns satisfy Billinge's criteria \cite{billinge2024materials} for a generalized materials gene and extend the materials genome concept to spatially heterogeneous crystalline materials. We propose the terms \textit{lattice gene} for the latent encoding of a single Kikuchi pattern and \textit{lattice genome} for the spatially resolved collection of lattice genes across a statistically representative area (or volume). We further demonstrate that the lattice genome enables microstructure analysis, including visualization, segmentation, and quantitative heterogeneity analysis in terms analogous to conventional orientation imaging and microstructure analysis.

\section{Lattice genome from encoding Kikuchi patterns}

First, we show that latent representations of Kikuchi patterns with variational autoencoders (or similar deep architectures) satisfy the criteria for materials genes proposed by Billinge \cite{billinge2024materials}. 

\begin{itemize}
\setlength{\itemsep}{0pt}

\item \textbf{Compactness}. The latent vector is a compact and low-dimensional representation of Kikuchi patterns (compared to raw images). 

\item \textbf{Experimental accessibility}. With a pre-trained encoder, the representation is experimentally accessible with a "forward pass" of Kikuchi patterns as measured with an EBSD detector. 

\item \textbf{Distance metric availability}. Kikuchi patterns are represented as points or vectors in a multidimensional latent space, where the distance between encoded patterns (\textit{e.g.}, \Cref{eqn:metrics}) reflects physical similarity of the probed lattice in terms of its crystallography, orientation, or defect content. This property is particularly advantageous compared to crystallographic representations, where a small structural distortion can discontinuously change the space group assignment \cite{billinge2024materials} and thus separate two nearly identical structures in the representation space. The continuous nature of the latent space avoids this limitation as structurally similar lattice states map to nearby points in the latent space.

\item \textbf{Informativeness and near-uniqueness}. The latent representation is sensitive to local microstructure at the length scale of the interaction volume for back-scattered electrons. Since Kikuchi patterns contain the lattice parameter, orientation, strain, and defect (\textit{e.g.}, dislocation) densities and the patterns can be reconstructed from the latent representation, one can expect that the latent representation efficiently encodes all this physically interpretable information. The ability to reconstruct the original Kikuchi pattern from the latent vector implies a near-unique mapping between the latent code and the underlying lattice state, within the bounds of the reconstruction error. In this context, while individual atomic positions and individual defects such as dislocations are beyond the reach of the EBSD resolution, the decoder trained in tandem with the encoder can serve as a "ribosome" capable of reconstructing the original Kikuchi pattern, from which the physical descriptors of the local crystalline structure can be inferred using published approaches \cite{wilkinson2006high,pantleon2008resolving,wang2023dislocation}. Calvat \textit{et al.}\ \cite{calvat2025learning} demonstrated that individual elements of the latent vector exhibit sensitivity to specific physical features such as crystallographic orientation, dislocation cell structures, lattice rotation domains, and surface integrity. This finding indicates that, although the latent representation entangles multiple quantities, it preserves physically meaningful information across its dimensions.

\end{itemize}

Based on these criteria, we propose the notion of lattice genes that carry essential information on the local crystalline structure of the material through latent representation of Kikuchi patterns. Since individual lattice genes carry information on the local structure at individual points in the microstructure probed during the measurement, an EBSD dataset covering a statistically representative area (or volume) of the microstructure can be thought of as the lattice genome for a given heterogeneous material in a given state. This terminology draws a direct parallel with biology. As a biological genome is the complete set of genes that together define an organism, the lattice genome is the complete spatial collection of lattice genes that together define the microstructural state of a heterogeneous material. Each lattice gene encodes the local crystalline structure at a single spatial point; the lattice genome captures how these local states are arranged across the material and thus encodes the spatial heterogeneity that governs macroscopic properties.

While latent representation encodes all the information contained in a Kikuchi pattern in a nearly lossless manner (within the reconstruction error), interpretation of individual elements in the lattice genes and their relation to familiar metallurgical descriptions can be non-trivial. In this context, the next question is whether the lattice genome can enable microstructure analysis similar to conventional EBSD workflows. To address this question, below we present three approaches to microstructure analysis using lattice genes, each mirroring a familiar task in conventional EBSD workflows: (i)  visualization of microstructure through low-dimensional projection of lattice genes, analogous to orientation imaging; (ii) segmentation of microstructure domains, analogous to grain reconstruction; and (iii) quantification of microstructure heterogeneity, analogous to kernel average misorientation and grain orientation spread. 

\begin{figure*}[t]
  \centering
  \includegraphics[width=\linewidth]{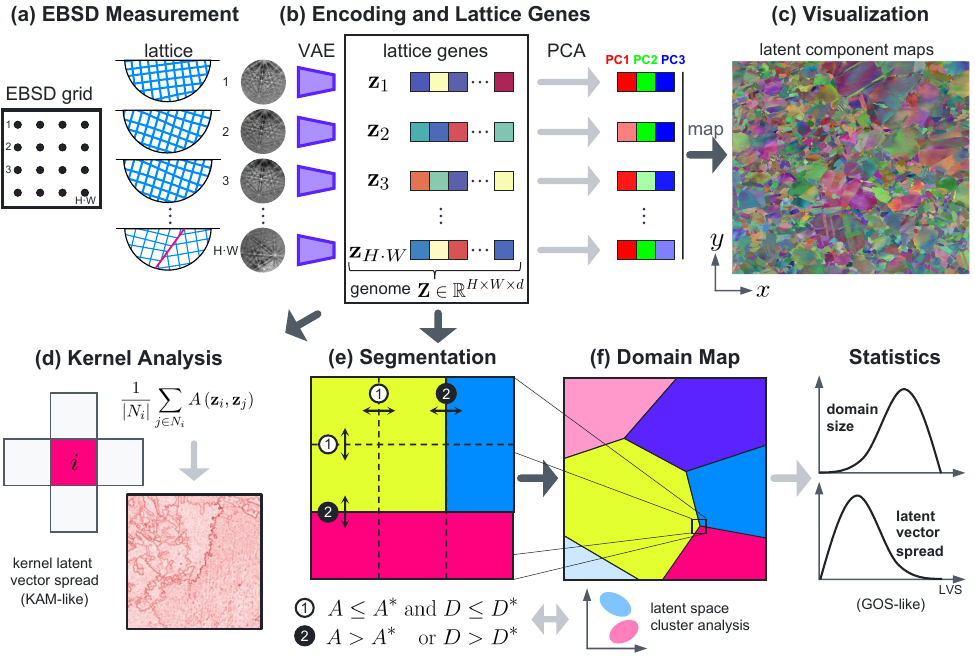}
    \caption{Overview of the present study, including the illustration of how the lattice in an interaction volume is probed during EBSD measurement at every grid point with Kikuchi diffraction. Encoding Kikuchi patterns results in lattice genes $\mathbf{z}$, that collectively make up a lattice genome, $\mathbf{Z}$, for a given microstructure, which enables microstructure imaging and analysis based on nearly lossless representation of the diffraction patterns.} 
  \label{fig:workflow}
\end{figure*}

\section{Microstructure analysis with lattice genome}
\label{sec:analysis}

Having established the conceptual basis for lattice genes and the lattice genome, we now describe a practical framework for microstructure analysis using these representations. The framework consists of three stages (\Cref{fig:workflow}): (i) encoding Kikuchi patterns into latent vectors using a pre-trained VAE, (ii) visualization through dimensionality reduction of the latent vectors with PCA, and (iii) segmentation of microstructure domains directly in the high-dimensional latent space with the optional steps of local heterogeneity and domain analysis. 

\textbf{Encoding}. Consider an $H\times W$ EBSD measurement grid with a Kikuchi pattern recorded at every grid point (\Cref{fig:workflow}(a)). These patterns are input as single-channel images into a pre-trained encoder \cite{calvat2025learning}, which represents each pattern as a $d$-dimensional latent vector, or lattice gene, $\mathbf{z}_i \in \mathbb{R}^d$. Encoding all patterns from the measurement (\Cref{fig:workflow}(b)) results in a $H\times W$ array of lattice genes which collectively constitute the lattice genome of the material, 
$\mathbf{Z} \in \mathbb{R}^{H\times W\times d}$. 

\textbf{Latent component maps}. For visual inspection of the lattice genome, we use PCA to reduce the $d$-dimensional latent vectors to three principal components (PC1, PC2, and PC3). For each principal component, the pixel-wise scores are normalized to the range $[0, 255]$ and assigned to the red (PC1), green (PC2), and blue (PC3) colors. Using the spatial coordinates of each $\mathbf{z}_i$, the RGB values are mapped to the EBSD grid and visualized as a regular RGB image (see \Cref{fig:workflow}(c)). This approach provides visually interpretable maps of the lattice genome while preserving as much variance of the original multidimensional lattice genes as possible within the three-channel constraint of vision and RGB displays. We refer to these images as \textit{latent component maps} to distinguish them from conventional orientation maps. Since PCA maximizes variance, high-contrast dimensions can dominate the leading principal components while subtler heterogeneities encoded in low-contrast dimensions are suppressed. To mitigate this effect, we additionally tested a pre-PCA scaling step that equalizes the contrast range across all latent dimensions (detailed in \Cref{sec:component-maps}).

\textbf{Segmentation of microstructural domains}. While classical grain analysis relies on thresholding local misorientations \cite{bachmann2011grain}, lattice genome representation of the microstructure requires new segmentation strategies operating on higher-dimensional representations corresponding to the lattice genes. In this study, we approach microstructure segmentation using two metrics: (i) the Euclidean distance between lattice genes as points in the latent space and (ii) the angle between lattice genes as vectors in the latent space. The distance, $D$, and the angle, $A$, are calculated for a pair of neighboring EBSD pixels, $i$ and $j$, as follows:

\begin{subequations}
\label{eqn:metrics}
\begin{align}
A_{i,j} &= \arccos\left ( \frac{\mathbf{z}_i\cdot\mathbf{z}_j}{\|\mathbf{z}_{i}\| |\mathbf{z}_{j}\|} \right ), \\
D_{i,j} &= \|\mathbf{z}_{i} - \mathbf{z}_{j}\|, 
\end{align}
\end{subequations}

\noindent where $\|\cdot\|$ is the $L_2$ norm. Equipped with these definitions, we compute $A$ and $D$ metrics for each pixel and two of its neighbors (down and right, assuming square grid), and compare the values against thresholds ($A^\ast$ and $D^\ast$) fixed for a given microstructure. If both metrics are below their thresholds (condition 1 in \Cref{fig:workflow}(e)), the neighbor pixel ($j$) is merged with the current pixel ($i$) into the same microstructural \textit{domain}. If, however, at least one metric is above its threshold (condition 2 in \Cref{fig:workflow}(e)), the algorithm will assign these pixels to different domains. Progressively scanning over the map row by row (from left to right) with pixel-wise merging and separation will constitute a segmented microstructure with domains assigned unique IDs (\Cref{fig:workflow}(f)). Since such segmentation does not necessarily lead to a \textit{grain} map defined by misorientations, we refer to the resulting constituents as \textit{domains}. 


The distance and angle thresholds ($A^\ast$, $D^\ast$) serve a similar role as misorientation threshold in conventional grain segmentation. However, while common values (\textit{e.g.}, \SI{5}{\degree} or \SI{10}{\degree}) are known for misorientation, thresholds for microstructure segmentation in lattice gene representation need to be established. In this study, we search for optimal thresholds by evaluating clusters of lattice genes in the latent space obtained for candidate thresholds. Specifically, we analyze clusters in terms of their total number, $K$ and their "compactness" quantified by within-cluster sum of squares (WCSS) \cite{macqueen1967multivariate}:

\begin{equation}
\text{WCSS} = \sum_{k=1}^{K} \sum_{\mathbf{z}_i \in C_k} \|\mathbf{z}_i - \boldsymbol{\mu}_k\|^2, 
\end{equation}

\noindent where $\mathbf{z}_i$ is the $i^{\mathrm{th}}$ lattice gene, $C_k$ is the set of lattice genes in a cluster $k$ with a centroid $\boldsymbol{\mu}_k$.

\begin{figure*}[h]
  \centering
  \includegraphics[width=\linewidth]{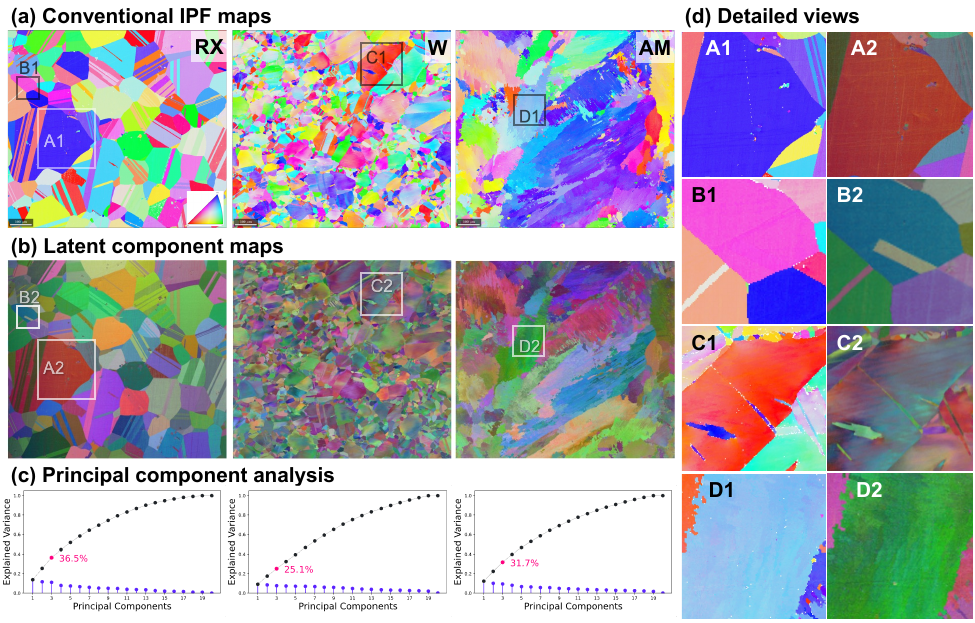}
  \caption{Microstructure visualization using (a) conventional inverse pole figure maps (IPF) and (b) new latent component maps. The IPF maps refer to the sample $x$ axis, the latent component maps assign RGB colors to the first three principal components of the 127-dimensional lattice genes. The three components capture 25 to \SI{37}{\percent} variance in these maps as seen in (c) explained variance plot. Latent component maps resolve microstructure details unseen in the IPF maps as shown in (d) detailed, close-up views of the microstructures.}
  \label{fig:PCA}
\end{figure*}

High thresholds $A^\ast$ and $D^\ast$ result in a small number of large domains with high internal heterogeneity and large WCSS. Tightening the thresholds results in a rapid decrease of WCSS as domains get split (see Fig.\ S2). At a certain point, further tightening yields only marginal WCSS reductions because the segmentation has resolved the meaningful microstructural boundaries and additional splits fragment domains that are already uniform. This transition produces a characteristic elbow in the WCSS curve (plotted against the number of clusters, $K$, see Fig.\ S2). We expect the elbow region corresponds to an optimal segmentation and thus target it when identifying the $A^\ast$ and $D^\ast$ thresholds.

Since distributions of $A$ and $D$ metrics vary across different microstructures  (see Fig.\ S1) with distinct grain sizes, textures, and defect content, absolute thresholds are not universal and thus do not necessarily transfer from one EBSD dataset to another. To ensure generality, we parameterize the thresholds in terms of percentiles of the $A$ and $D$ distributions computed from all neighboring pixel pairs in each map. Specifically, we define a single percentile parameter, $p$, and set $A^\ast$ and $D^\ast$ thresholds to the $p$-th percentile of their respective distributions. This strategy relates the thresholds to the intrinsic statistical characteristics of each microstructure. 

Our preliminary tests show that scanning $p$ from 95th (loose) to 60th (strict) percentile leads to efficient identification of optimal segmentation. Thresholds above the 95th percentile overlook obvious boundaries, while those below the 60th percentile produce excessive fragmentation. Since identifying the elbow from the full and fine WCSS curve can be computationally intense for large maps, we adopt an early stopping scheme \cite{yao2007early, prechelt2002early}, which terminates the search once WCSS reductions become marginal. Early stopping combined with screening from loose thresholds (where segmentation is faster) makes the optimal threshold search computationally inexpensive. The detailed parameters of the early-stopping scheme, full distributions of $A$ and $D$ of three cases, and the corresponding elbow plots for WCSS are provided in the Supplementary Material.

\section{Case study: Ni-base superalloys in diverse microstructure states}
\label{sec:results}

We demonstrate the proposed framework on two Ni-base superalloys (Inconel 718 and Waspalloy) with the analysis of their microstructures in three states: (i) wrought IN718 after full recrystallization (labeled as \textbf{RX} hereafter); (ii) wrought Waspalloy after partial recrystallization (\textbf{W}); and (iii) as-built IN718 additive manufactured (\textbf{AM}) with direct energy deposition \cite{calvat2025kikuchipattern}. These states represent distinct microstructures with well-defined grains (RX), well-defined grains with intragranular (orientation and defect) variations (W), and less-defined grains with significant intragranular variations (AM). The EBSD data for these microstructures and their encodings were obtained by Calvat \textit{et al.} \cite{calvat2025kikuchipattern}, who gave detailed description of the alloys and their characterization, as well as latent encoding \cite{calvat2025learning}. Note that the original latent representation presented by Calvat \textit{et al.} \cite{calvat2025learning} contains 128 dimensions. In this study, we exclude one of the elements from all lattice genes because we found that it mostly captures effects related to measurement geometry, which show a variation along the horizontal direction. As a result, all lattice genes in this study have $d=127$. The compositions and the processing of these three alloys are briefly outlined in the Supplementary Material. 

\subsection{Latent component maps}
\label{sec:component-maps}

The latent-component visualization of the lattice genome captures the microstructural heterogeneity similar to the classical orientation maps (\Cref{fig:PCA}). The spatial heterogeneity is captured in all three cases with both well- and less-defined grains and even though the first three principal components capture no more than \SI{37}{\percent} of variance in lattice genes of each individual map (\Cref{fig:PCA}(c)). In addition, the latent component maps resolve some microstructural features barely visible or less evident in the conventional inverse pole figure maps (\Cref{fig:PCA}(d)): \textit{e.g.}, plastic deformation artifacts from sample preparation (A1 vs.\ A2) or even additional domain/grain (B1 vs.\ B2) in the RX case; smooth intragranular gradients likely associated with orientation variations in the W case (C1 vs.\ C2), and intragranular substructure likely associated with dislocation cells in the AM case (D1 vs.\ D2). 

\begin{figure*}[t]
  \centering
  \includegraphics[width=\linewidth]{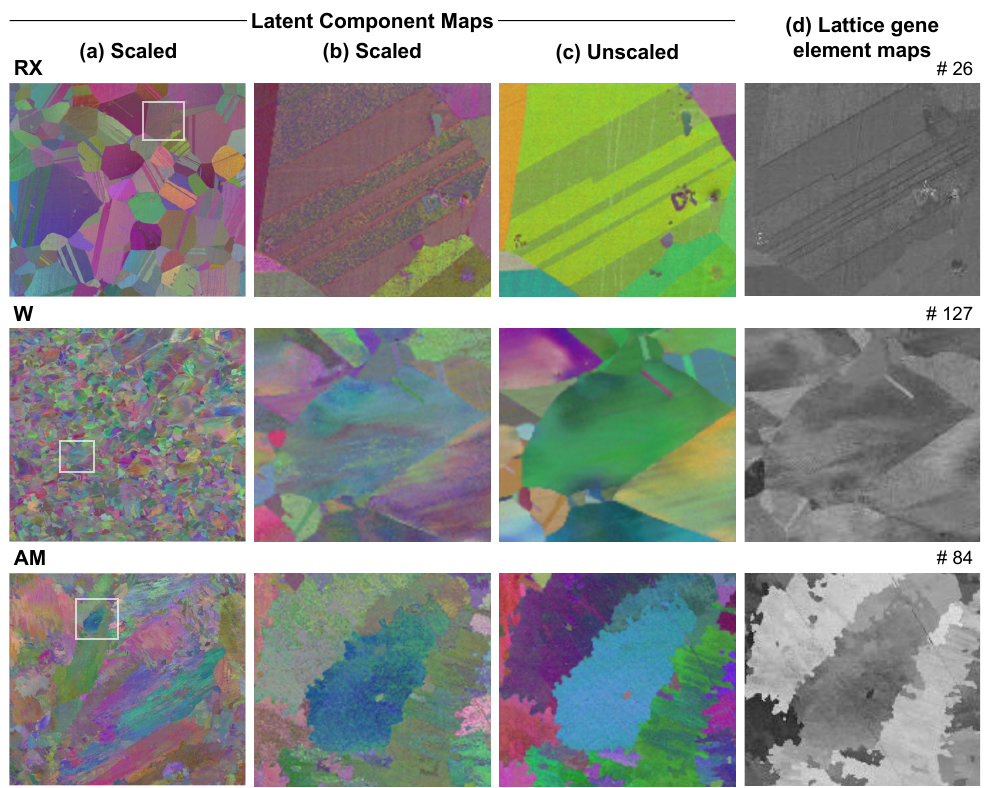}
  \caption{Scaled latent component maps (a,b) help reveal subtle microstructure details overlooked in unscaled latent component maps (c) that are only visible in the maps of individual lattice gene dimensions (d).}
  \label{fig:Norm_PCA}
\end{figure*}

Although lattice genes encode richer information than crystallographic orientations alone \cite{calvat2025learning}, the latent component maps do not automatically reveal all of this information. The three-component PCA projection is inherently lossy because the explained variance analysis shows that 20 components are needed to explain nearly \SI{100}{\percent} in all three microstructures (\Cref{fig:PCA}(c)). Further, since PCA prioritizes directions of maximum variance, high-contrast features dominate the leading components \cite{latypov2017data}. As a result, the PCA-based maps may overlook subtler heterogeneities, such as intragranular gradients or gradual defect variations. We found that scaling individual dimensions of the lattice genes can help resolve these subtle, ``low-contrast'' microstructure details. Specifically, for each latent dimension, the values across all $H \times W$ pixels are rescaled so that the 1st and 99th percentiles map to 0 and 1. This equalizes the contrast range across all 127 dimensions before PCA and prevents high-contrast features from dominating the leading principal components. Such pre-PCA scaling can reveal subtle details without the need to inspect individual elements of the full 127-dimensional lattice genes. Scaled latent component maps (\Cref{fig:Norm_PCA}(a,b)) reveal such features as an annealing twin and intragranular gradients unseen in the unscaled latent component maps (\Cref{fig:Norm_PCA}(c)). These features can be otherwise found only in the maps of individual latent gene dimensions (\Cref{fig:Norm_PCA}(d)). 

\begin{figure*}[t]
  \centering
  \includegraphics[width=\linewidth]{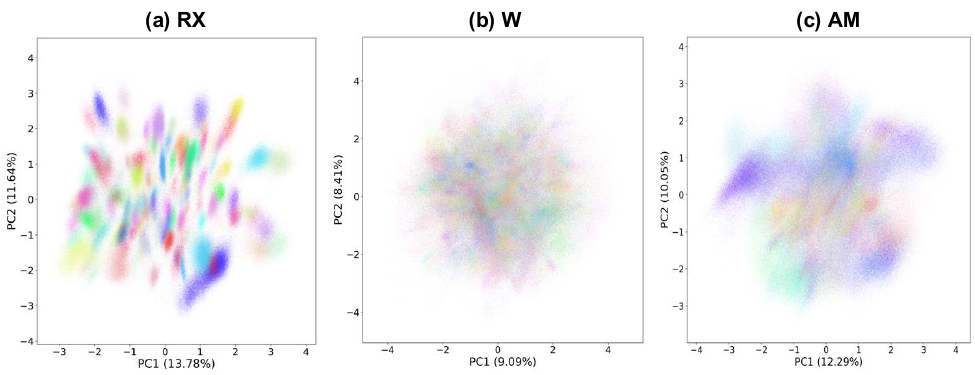}
  \caption{Visualization of 127-dimensional lattice genes in two dimensions using PCA for (a) RX, (b) W, and (c) AM microstructures.}
  \label{fig:projection}
\end{figure*}

We complement the latent component maps with the visualization of all three lattice genomes in the two-dimensional space in terms of the first two principal components (\Cref{fig:projection}). The low-dimensional visualization shows clusters, especially in the RX case with well-defined grains (\Cref{fig:projection}a). This result supports the feasibility of microstructure segmentation with the aid of clustering in the high-dimensional space of lattice genes presented in the next section. \\

\subsection{Microstructure segmentation}

We applied the segmentation framework proposed in \Cref{sec:analysis} to the three microstructures with thresholds determined by the percentile screening procedure with early stopping. The thresholds were identified corresponding to the 76th, 80th, and 73rd percentiles 
of the $A$ and $D$ distributions (see Fig.\ S1) for the RX, W, and AM samples, respectively. These thresholds capture the elbow regions of the WCSS curves (see Fig.\ S2) expected to result in the optimal segmentation. Segmentation with both conventional and present methods dismissed grains consisting of fewer than six pixels. 

\begin{figure*}[t]
  \centering
  \includegraphics[width=\linewidth]{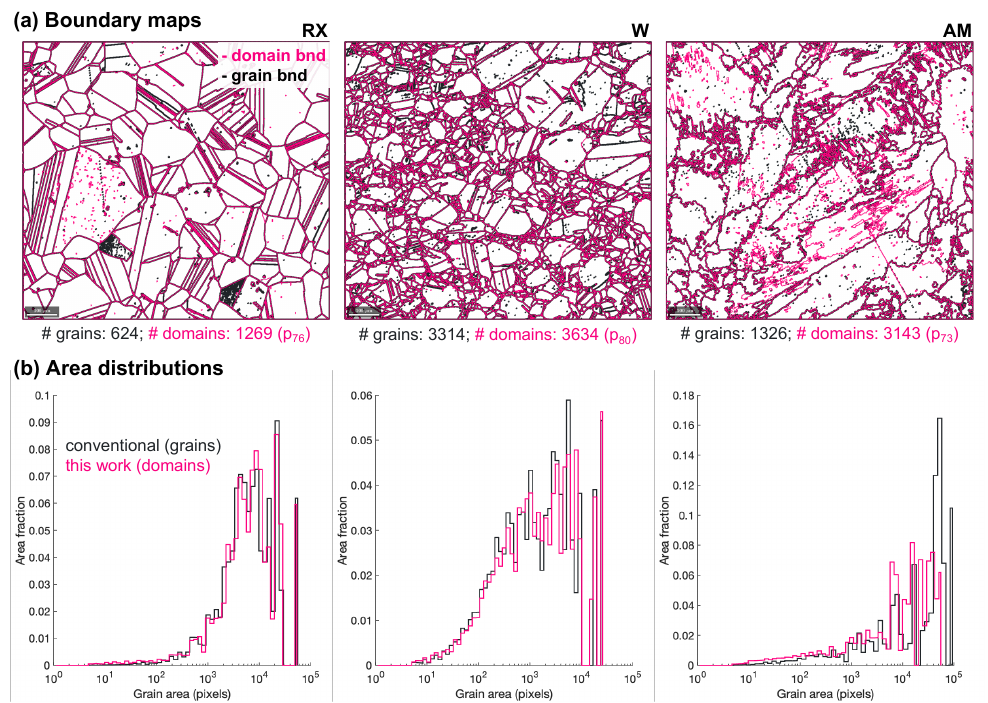}
  \caption{Segmentation of lattice genes produce domain maps largely matching grain maps as seen in (a) overlay boundary networks and (b) area distributions. The grain boundaries were obtained by \SI{5}{\degree} misorientation thresholding, while percentile-based procedure was used for domain boundaries with the resulting percentile ($p_{76}$, etc.) displayed under each map.}
  \label{fig:segmentation}
\end{figure*}

The resulting domain boundary networks largely coincide with the traditional grain boundary maps obtained by misorientation thresholding (\Cref{fig:segmentation}(a)). While the overlap is significant (especially for RX and W states), the boundary networks also exhibit differences. The domain segmentation is more sensitive than conventional \SI{5}{\degree} misorientation thresholding as seen in boundary segments identified exclusively in the domain network as well as in the number of grains vs.\ domains (624 vs.\ 1269 for RX, 3314 vs.\ 3634 for W, and 1326 vs.\ 3143 for AM). The difference goes beyond new domain boundaries unobserved in grain boundary maps. The lattice genome segmentation also leaves out some grain boundaries: those of small grains in AM and W conditions but also some boundaries most likely associated with indexing problems (dark patches in RX map in \Cref{fig:segmentation}(a)) as well as a few regular grain boundaries. 

The segmentation of domain and grain boundary networks also show quantitative agreement in terms of size distributions (\Cref{fig:segmentation}(b)). The grain and domain size distributions (in terms of areas) closely match with the exception of the AM microstructure. The difference for the AM case comes from the many domain boundaries not present in the grain segmentation. 

\subsection{Quantifying of microstructural heterogeneity}

Beyond segmentation, lattice gene representation enables quantification of microstructural heterogeneity at multiple scales. At the pixel level, conventional EBSD analysis commonly uses the kernel average misorientation (KAM), defined as the average misorientation angle between a pixel and its nearest neighbors~\cite{schwartz2009electron}. By direct analogy, we define the kernel latent vector spread (KLVS) as the average cosine angle between a lattice gene at a pixel, $\mathbf{z}_i$, and the lattice genes of its nearest neighbors, $\mathbf{z}_j$:

\begin{equation}
\mathrm{KLVS}_{i} = \frac{1}{|N_{i}|} \sum_{j \in N_{i}} A\left( \mathbf{z}_{i}, \mathbf{z}_{j} \right),
\label{eqn:kam}
\end{equation}

\noindent where $N_{i}$ is the set of nearest neighbors of the pixel $i$ with cardinality $|N_{i}|$.

\begin{figure*}[t]
  \centering
  \includegraphics[width=\linewidth]{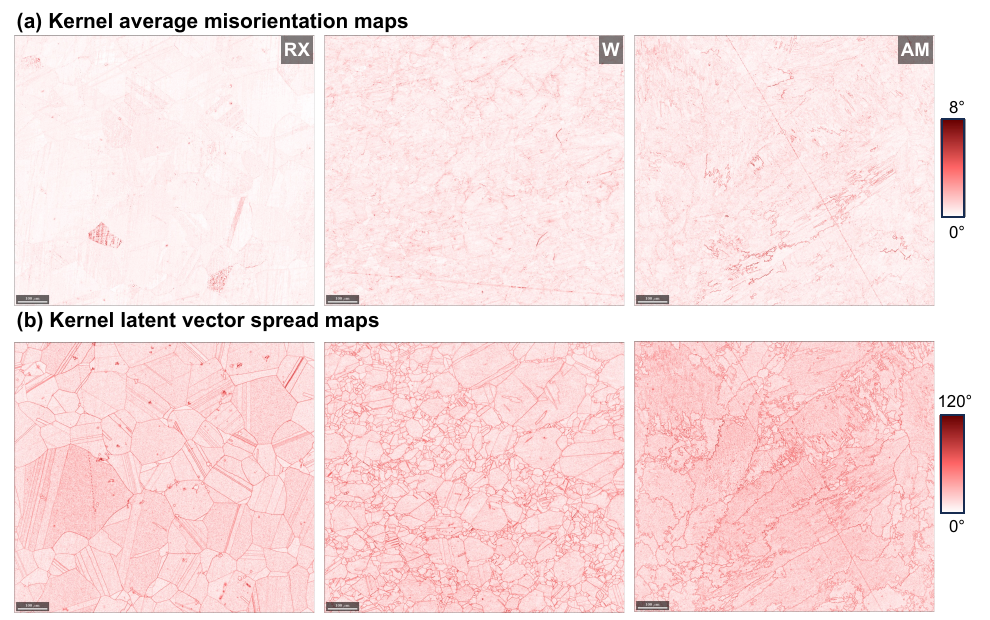}
  \caption{Classical (a) kernel average misorientation and new (b) kernel latent vector spread maps showing local heterogeneity in the three microstructures. Note KAM indexing artifacts in three grains in the RX sample and a scratch in the W sample -- both absent in the corresponding KLVS maps.} 
  \label{fig:kam}
\end{figure*}

\Cref{fig:kam} compares conventional KAM and KLVS maps for the three microstructures. Both metrics reproduce the expected qualitative trends: concentration of high values along boundaries in RX, elevated values throughout W reflecting stored deformation, and patches of elevated values in AM. KLVS, however, produces higher visual contrast for real microstructural features. In KAM maps, the dynamic range is impacted by elevated values that originate from artifacts rather than from the underlying microstructure. As example, clusters of high KAM values correspond to three grains with problematic indexing in the RX sample (\Cref{fig:kam}(a), lower half). The same regions are clean in the KLVS map (\Cref{fig:kam}(b)). Another example is a polishing scratch traversing the W microstructure in the KAM map that is less prominent in the KLVS map. KLVS reduces these artifacts as it operates on the encoded pattern instead of extracted orientations. Perturbations that disturb band detection without altering the overall diffraction signal have limited effect on the lattice gene.

The gain in contrast for real features is particularly evident at grain boundaries, where the lattice gene encodes the superposition of Kikuchi patterns from two grains and naturally differs from the single-lattice genes of either neighbor. KLVS also resolves intragranular substructure within larger grains in the AM sample (\Cref{fig:kam}(b)) that is not visible in the corresponding KAM. Similar features have previously been associated with dislocation cellular structures and local lattice rotation domains in VAE-encoded AM material~\cite{calvat2025learning}.

As seen in the color ranges in (\Cref{fig:kam}), absolute KLVS values (\SIrange{0}{120}{\degree}) are not directly comparable to KAM (\SIrange{0}{8}{\degree}) because latent-space angles are not reduced by crystal symmetry. We further tested a distance-based version of KLVS, where the Euclidean distance between lattice genes is used instead of the cosine angle (see Supplementary Materials). The results are similar to the angle-based KLVS (see Fig.\ S3), which suggests that both angle and distance in the latent space serve as consistent descriptors of local microstructural variation. 

\begin{figure*}[ht!]
  \centering
  \includegraphics[width=\linewidth]{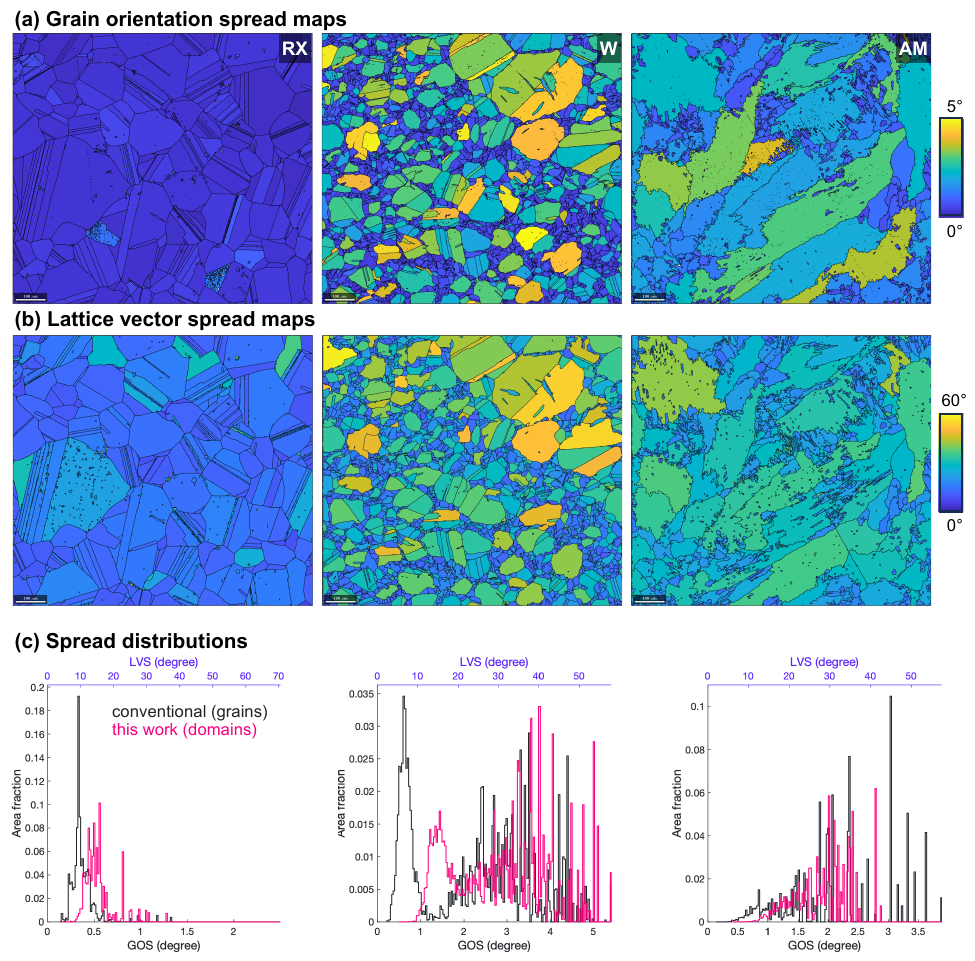}
  \caption{Classical grain orientation spread and new latent vector spread show qualitative similarities in terms of (a,b) maps and (c) distributions. LVS maps are shown for the domains segmented as shown in \Cref{fig:segmentation}.}
  \label{fig:gos}
\end{figure*}

Beyond pixel-level analysis with KLVS, lattice genes can also quantify grain-level heterogeneity. In classical orientation-based EBSD analysis, grain orientation spread (GOS) is commonly used towards this end. GOS is defined as the average misorientation angle between each pixel orientation and the reference (mean) orientation of its corresponding grain \cite{schwartz2009electron, wright2011review}. With lattice genes, we introduce latent vector spread (LVS) as a high-dimensional analog of GOS:

\begin{equation}
\mathrm{LVS} = \frac{1}{N} \sum_{i} A\left( \boldsymbol{\mu}, \mathbf{z}_{i} \right),
\label{eqn:lvs}
\end{equation}

\noindent which averages the cosine angles, $A$, between the lattice gene at a pixel, $\mathbf{z}_i$, and the centroid of the lattice gene cluster in the latent space, $\boldsymbol{\mu}$, corresponding to the current microstructure domain containing $N$ pixels. Calculation of LVS for each domain allows for spatially resolved visualization (\Cref{fig:gos}(b)) and quantitative analysis in terms of distributions (\Cref{fig:gos}(c)).

The LVS maps exhibit qualitative trends very similar to conventional GOS maps (\Cref{fig:gos}(a,b)): low spread in recrystallized grains (RX), high spread in worked but well-defined grains (W), and mid-range spread in the microstructure with long-range heterogeneities and less defined grains (AM). Since LVS is derived from angles in the high-dimensional latent space, its magnitude is distinct from GOS even though it is still expressed in degrees. The conceptual difference in the calculation combined with the difference in segmentation leads to distinct distributions of GOS vs.\ LVS (\Cref{fig:gos}(c)). The area-weighted GOS distributions are most distinct for the AM microstructure, yet display similarities for the RX and W cases where similar but shifted peaks are observed. 

\section{Discussion}

The lattice genome framework offers a more complete representation of EBSD data than conventional descriptors by preserving nearly all information contained in Kikuchi diffraction patterns. Our results demonstrate that familiar microstructure analysis tasks (visualization, segmentation, and quantitative assessment of heterogeneity) can be carried out using lattice genes without extracting explicit crystallographic quantities. To this end, we introduced latent component maps as analogs of orientation maps, domain segmentation based on distance and angle metrics in the latent space, and latent vector spreads as high-dimensional counterparts to GOS and KAM. In all three studied microstructures, these tools produced results broadly consistent with conventional EBSD analysis while, in some cases, revealing finer heterogeneities attributable to the richer information content of the latent representation.

This more complete representation, however, comes at the cost of reduced interpretability compared to conventional descriptors. IPF maps (\Cref{fig:PCA}(a)) use a universal color scheme tied directly to crystallographic orientations: a given color carries the same physical meaning across any map of a material with the same crystal symmetry. Latent component maps (\Cref{fig:PCA}(b)), by contrast, are dataset-dependent. All three EBSD datasets were processed with the same encoder, which leads to a consistent lattice genome representation in the shared latent space. However, the PCA was performed on individual maps so that the colors are not transferable from map to map, unless a single PCA is carried out on a combined dataset. Latent component maps therefore serve primarily as tools for revealing spatial heterogeneity \textit{within} a given microstructure considering the entirety of Kikuchi patterns. 

A related limitation concerns the crystal symmetry. In conventional EBSD analysis, misorientations are computed by explicitly accounting for the symmetry operators of the crystal system, which ensures that physically equivalent orientations are recognized. Crystal symmetry is almost certainly encoded within lattice genes (the encoder learns from Kikuchi patterns that inherently reflect the point group) but this encoding is implicit and entangled across dimensions rather than explicitly enforced. This distinction has practical consequences for the latent vector spread analysis. Unlike GOS and KAM, which apply symmetry to minimize misorientation, LVS and KLVS compute angles between raw latent vectors without such reduction (\Cref{eqn:kam,eqn:lvs}). This explains why KLVS and LVS values are systematically larger than KAM and GOS as well as why the GOS and LVS distributions show similar trends but do not match quantitatively (\Cref{fig:kam,fig:gos}).

A further caveat on the lattice genes considered here concerns information contained in Kikuchi patterns that is not intrinsic to the material. Recorded diffraction patterns inherently contain the EBSD measurement conditions such as detector geometry and pattern-center position, background variations, and acquisition noise \cite{britton2016tutorial}. Since the encoders are trained to reconstruct the full pattern, these non-material features become embedded in the lattice gene alongside the lattice information. The present lattice genes are therefore sensitive to the microstructure \textit{and} the specific measurement conditions and can complicate comparisons across datasets acquired with different detectors, geometries, or settings, which however can be addressed with encoders trained on larger and more diverse datasets as discussed below.

A related consideration arises when the electron beam interaction volume spans multiple lattices, which can occur at interfaces, \textit{e.g.}, grain or twin boundaries \cite{humphreys2001review,zaefferer2007formation}. The resulting Kikuchi pattern is a superposition of contributions from the different lattices in the interaction volume \cite{zhu2020ebsd,calvat2025learning}, and the corresponding lattice gene therefore encodes this overlap rather than a single lattice state. The effect is two-fold. On one hand, lattice genes at such locations are harder to interpret in terms of a single well-defined lattice. On the other hand, the encoding is sensitive to such overlaps unlike conventional indexing, which assumes a single pattern per pixel. This sensitivity, previously noted by Calvat et al.~\cite{calvat2025learning}, contributes to the finer boundary networks observed in our domain segmentation compared with conventional grain reconstruction (\Cref{fig:segmentation}).

Despite these limitations, the lattice genome framework offers practical advantages. First, it operates entirely on the raw diffraction patterns without Hough-based indexing, dictionary matching, or any other preprocessing step. This makes it applicable to patterns that are difficult to index conventionally, such as those from heavily deformed or multiphase regions \cite{wright2015introduction}. Second, segmentation approach introduced here uses a single percentile parameter to produce domain maps across microstructures with very different statistical characteristics, which avoids the need for manual case-by-case threshold calibration due to non-universal nature of thresholds ($A^\ast$ and $D^\ast$, see details in \Cref{sec:analysis}). Third, because the full Kikuchi pattern information is preserved in the lattice gene, the same encoded dataset can serve as input for diverse downstream tasks (\textit{e.g.}, microstructure analysis or even property prediction) without reprocessing the raw patterns.

Further, some of the interpretability and entanglement challenges are not inherent in the lattice gene concept and can be addressed through the development of improved, physically constrained encoders. Rotation-invariant architectures \cite{burgess2024orientation} or conditional VAEs trained with crystallographic supervision could disentangle \cite{wang2024desiderata,locatello2019challenging} physical quantities (\textit{e.g.}, orientation and defect content) or make crystal symmetry explicit in the latent representation. Finally, special and extended training strategies for the encoder (large datasets, data augmentation, contrastive learning) could further make the encoding invariant to the instrument or measurement settings (mentioned above) and thus distill the lattice gene to intrinsically microstructural content. Indeed, by collecting and including in training increasingly large and diverse diffraction patterns from different alloys of different crystal symmetries, we expect that a unique encoding with universal distance (and thus similarity) metric in the latent space can be achieved. 

As encoder architectures evolve, the concepts and analysis tools presented here will remain applicable. The lattice gene and lattice genome are defined by encoding Kikuchi patterns into compact latent vectors independent of the specifics of any particular encoder architecture. Similarly, the microstructure analysis tools we introduced (latent component maps, segmentation, and heterogeneity quantification) can operate on any high-dimensional representations and are therefore applicable to lattice genes produced by any encoder.

While only a few of representative analyses have been demonstrated in this study, lattice genome representation allows many other well-established and more advanced microstructure workflows. For example, the domain maps obtained with proposed segmentation strategy (\Cref{fig:segmentation}) can be readily analyzed with angularly- or spatially-resolved chord length distributions \cite{latypov2018application,whitman2025sr}. Similar to LVS presented here, high-dimensional analogs to other orientation-based metrics such as grain-average misorientation \cite{wright2011review,field2005analysis}, grain reference orientation distribution \cite{schwartz2009electron}, or $n$-point correlation functions \cite{paulson2017reduced,adams1998mesostructure} can be introduced for the lattice genome representation.

Extending the lattice genome to multiphase materials, where Kikuchi patterns from different crystal structures coexist, would test the generalization of the concept beyond single-phase polycrystals. Coupling lattice genes with complementary chemical information (\textit{e.g.}, from energy-dispersive X-ray spectroscopy) could enable unified encoding of microstructures with both structural and chemical heterogeneity. In either case, the lattice genome provides a compact, spatially resolved, and information-rich representation of microstructure that is naturally suited as input for machine learning models linking microstructure to properties or processing. This direction complements microstructure informatics efforts in the field \cite{richter2025microstructure,thome2026neural}. While published approaches rely on hand-crafted descriptors of specific microstructural features, the lattice genome provides a continuous, learned representation of the underlying diffraction signal. In this context, the lattice genome may offer a practical realization of the MGI vision for the mesoscale of defect-dominated and highly heterogeneous crystalline materials.

\section{Conclusion}

In this work, we introduced the notions of lattice gene and lattice genome for heterogeneous crystalline materials. A lattice gene obtained by encoding a Kikuchi diffraction pattern satisfies the criteria for a generalized materials gene proposed by Billinge \cite{billinge2024materials}: compactness, experimental accessibility, a distance-respecting metric, and sufficient information content for near-lossless reconstruction of the original diffraction signal. The spatially resolved collection of lattice genes across a representative area (or volume) constitutes the lattice genome, which extends the materials genome concept to heterogeneous crystalline materials at the mesoscale.

We demonstrated that the lattice genome enables microstructure analysis analogous to conventional EBSD workflows yet capable of resolving finer microstructural details. Latent component maps constructed from PCA of lattice genes reveal grain-scale and intragrain heterogeneities comparable to orientation maps  while revealing more subtle structural variations. Microstructure segmentation into domains based on distance and angle metrics in the high-dimensional latent space produces boundary networks consistent with conventional grain boundary maps. Latent vector spreads, introduced as high-dimensional analogs of grain orientation spread and kernel average misorientation, recover qualitatively similar trends in microstructure heterogeneity across all additively manufactured and recrystallized states in the studied Ni-base superalloys, while offering better contrast. All developed tools operate directly on encoded Kikuchi diffraction data without pattern indexing. Acting directly on high-dimensional vectors independent of the encoder specifics, they will remain applicable as encoder designs evolve.

\section*{Data and code availability}

The raw EBSD datasets corresponding to the case study are publicly available from Calvat \textit{et al.}\ \cite{calvat2025kikuchipattern}. Codes for genomic microstructure analysis will be made public upon acceptance of the manuscript for publication. 

\section*{Acknowledgments}

MIL acknowledges the support by the National Science Foundation under Award No.\ 2441813. M.C.\ and J.C.S.\ are grateful for financial support from the Defense Advanced Research Projects Agency (DARPA -- HR001124C0394). The authors further thank Carpenter Technology for providing the wrought IN718 and Waspaloy used in this study. 

\begin{figure*}[t]
  \centering
  \includegraphics[width=\linewidth]{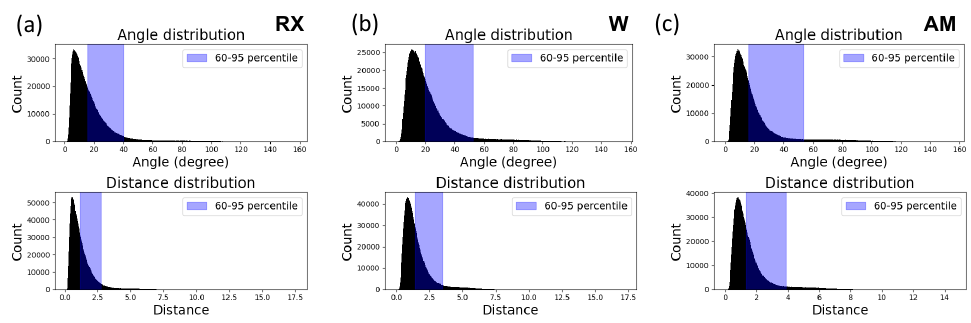}
  \caption{Distributions of $A$ and $D$ for neighboring pixels. The shaded regions shows the percentile range used for threshold selection. }
  \label{fig:distribution}
\end{figure*}

\section*{Supplementary Material}

\subsection*{Materials: composition and processing}

\paragraph{RX microstructure: Inconel 718 wrought and recrystallized}

\begin{itemize}
\setlength{\itemsep}{0pt}

\item \textbf{Composition} (wt.\%): Ni (base) – 0.56\% Al – 17.31\% Fe – 0.14\% Co – 17.97\% Cr – 5.4\% Nb+Ta – 1.00\% Ti – 0.023\% C – 0.0062\% N

\item \textbf{Processing}: \SI{1050}{\degreeCelsius} during 30 minutes to produce uniform texture, followed by 8 hours at \SI{720}{\degreeCelsius}, forming $\gamma$' and $\gamma$'' precipitates.

\end{itemize}

\paragraph{W microstructure: Waspalloy wrought and partially recrystallized}

\begin{itemize}
\setlength{\itemsep}{0pt}

\item \textbf{Composition} (wt.\%): 13.50\% Co - 19.50\% Cr - 4.25\% Mo - 2.00\% Fe - 1.40\% Al - 3.00\% Ti - 0.50\% Cu, 0.030\% P - 0.006\% B - 0.030\% S - 0.06\% Zr - 0.75\% Si - 1.00\% Mn - 0.070\% C

\item \textbf{Processing}: forged at high temperature to trigger a bimodal microstructure and heat-treated at \SI{1050}{\degreeCelsius} for 12 hours followed by 1 hour at \SI{760}{\degreeCelsius}.

\end{itemize}

\paragraph{AM microstructure: Inconel 718 additively manufactured}

\begin{itemize}
\setlength{\itemsep}{0pt}

\item \textbf{Composition} (wt.\%): Ni (base) – 0.45\% Al – 18.77\% Fe – 0.07\% Co – 18.88\% Cr – 5.08\% Nb – 0.96\% Ti – 0.036\% C – 0.02\% Cu - 0.04\% Mn - 0.08\% Si - 3.04\% Mo

\item \textbf{Processing}: Formalloy L2 Directed Energy Deposition utilizing a \SI{650}{\watt} Nuburu \SI{450}{\nano\meter} blue laser (with a spot size up to \SI{400}{\micro\meter}).

\end{itemize}





\subsection*{Distance and angle distributions in the latent space}

\Cref{fig:distribution} presents the distributions of cosine angles and Euclidean distances for neighboring pixels of three cases: RX, W, AM. The purple shaded region indicates the range  between 60th and 95th percentiles, within which the optimal threshold is identified.

\subsection*{Early-stopping scheme}

As $p$ gradually decreases from the 95th to 60th percentile with step of 1, the relative  reduction in within-cluster sum of squares (WCSS) at step $t$, $r(p_t)$ was tracked according to the expression:

\begin{equation}
r(p_t) = \frac{\mathrm{WCSS}(p_{t-1}) - \mathrm{WCSS}(p_t)}{\mathrm{WCSS}(p_0)},
\end{equation}

\noindent where $p_0$ denotes the loosest threshold. To determine saturation, a baseline was estimated based on the mean WCSS reduction of the first eight tightening steps. Early stopping was triggered when three consecutive steps all fell below \SI{10}{\percent} of this baseline, indicating diminishing structural refinement.

\begin{figure*}[t]
  \centering
  \includegraphics[width=\linewidth]{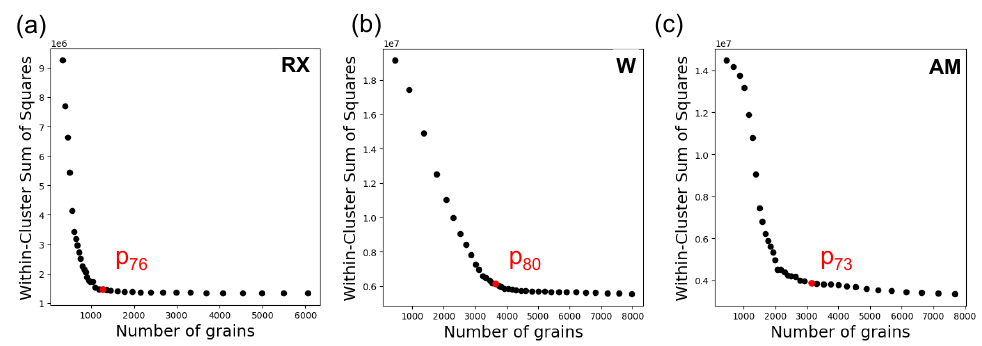}
  \caption{WCSS vs.\ the number of grains.}
  \label{fig:WCSS}
\end{figure*}

The final threshold (optimal $p^\ast$) was selected as the midpoint of the detected plateau window (three consecutive steps) to enhance robustness. This automated procedure enables efficient elbow detection without reliance on subjective visual inspection of the WCSS curve, which ensures objectivity and reproducibility. Furthermore, this procedure (unlike absolute threshold values) readily generalize to any EBSD dataset. Identical parameter settings were applied for all three cases. \Cref{fig:WCSS} shows the full elbow plots for WCSS as function of number of grains. The red dots indicate the optimal $p^\ast$ obtained from early-stopping scheme.

\subsection*{Distance-based kernel latent vector spread}

Analogous to angle-based KLVS presented in the main text, the distance-based KLVS (D-KLVS) is introduced as: 

\begin{equation}
\mathrm{D\text{-}KLVS}_{i} =
\frac{1}{|N_{i}|}
\sum_{j \in N_{i}}
\left\|
\mathbf{z}_{i}
-
\mathbf{z}_{j}
\right\|,
\label{eqn:kam}
\end{equation}

\noindent $\|\cdot\|$ is the $L_2$ norm, $N_{i}$ is the set of nearest neighbors of the pixel $i$ with cardinality $|N_{i}|$. 

\begin{figure*}[t]
  \centering
  \includegraphics[width=\linewidth]{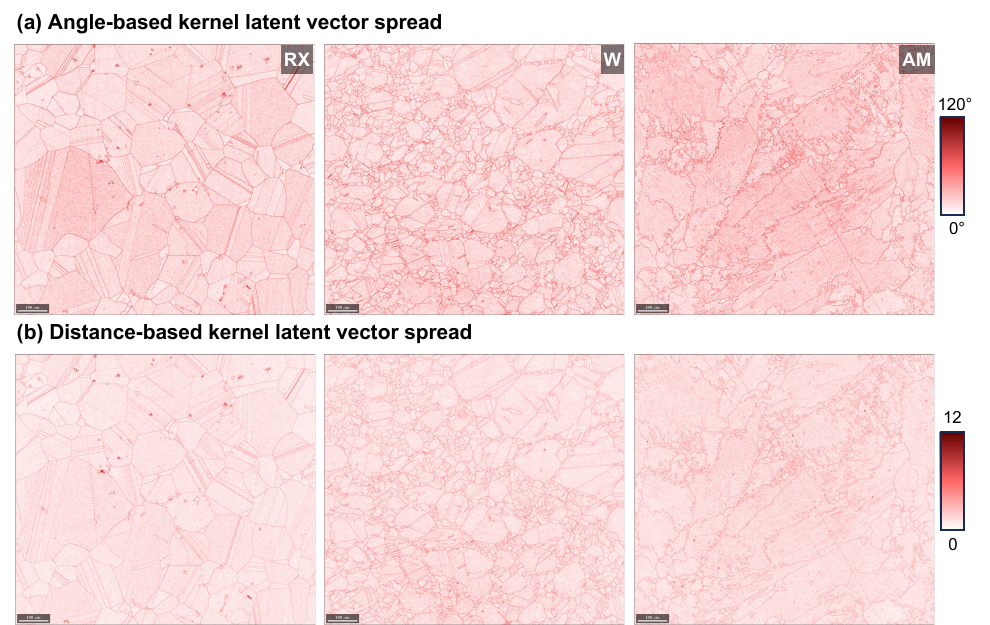}
  \caption{Kernel latent vector spread based on cosine angle (a) and Euclidean distance (b). }
  \label{fig:KLVS}
\end{figure*}

\Cref{fig:KLVS} presents the angle-based and distance-based KLVS maps for three microstructured studied here. The two sets of maps exhibit only minor differences. Their overall similarity indicates that both angular deviation and Euclidean distance in the latent space provide consistent descriptors of local microstructural heterogeneity.

\bibliography{references}
\end{document}